\documentclass[12pt]{article}

\def\lesssim{\mathrel{\hbox{\rlap{\hbox{\lower4pt\hbox{$\sim$}}}\hbox{$<$}}}}
\def\gtrsim{\mathrel{\hbox{\rlap{\hbox{\lower4pt\hbox{$\sim$}}}\hbox{$>$}}}}

\def \ppar{$~$ \\ \noindent}

\def\HI{H$^\circ$}
\def\lya{Ly$\alpha$}
\def\Vhc{$V_\mathrm{HC}$}

\newcommand{\aap}{{A\&A}~ }

\newcommand{\apjs}{{ApJS}~}

\usepackage{times}
\usepackage{epsfig}
\begin{document}
%\renewcommand{\baselineskip}{18pt}

%%%%%%%%%%%%%%%%%%%%%%%%%%%%%%%%%%%%%%%%%%%

\begin{center}
{ \bf \large Multi-Cycle HST Treasury Program for STIS: } \\  
{\bf \Large Mapping the Galactic Environment of the Sun} \\
{$ ~ $} \\
{ \large Priscilla C. Frisch\footnote{University of Chicago, Department of Astronomy and
Astrophysics, 5640 South Ellis Avenue, Chicago, IL 60637 USA; Phone:  773-702-0181; Email:  frisch@oddjob.uchicago.edu}} 
\end{center}

%%%%%%%%%%%%%%%%%%%%%%%%%%%%%%%%%%%%%%%%%%%%%%%%%%%%%%%%%%%%%
\section{Treasury Program Goal:  High S/N and Resolution Survey of Interstellar
Clouds within 20 pc of Sun }

\begin{figure}[!t]
\begin{center}
\flushleft
\includegraphics[scale=0.4]{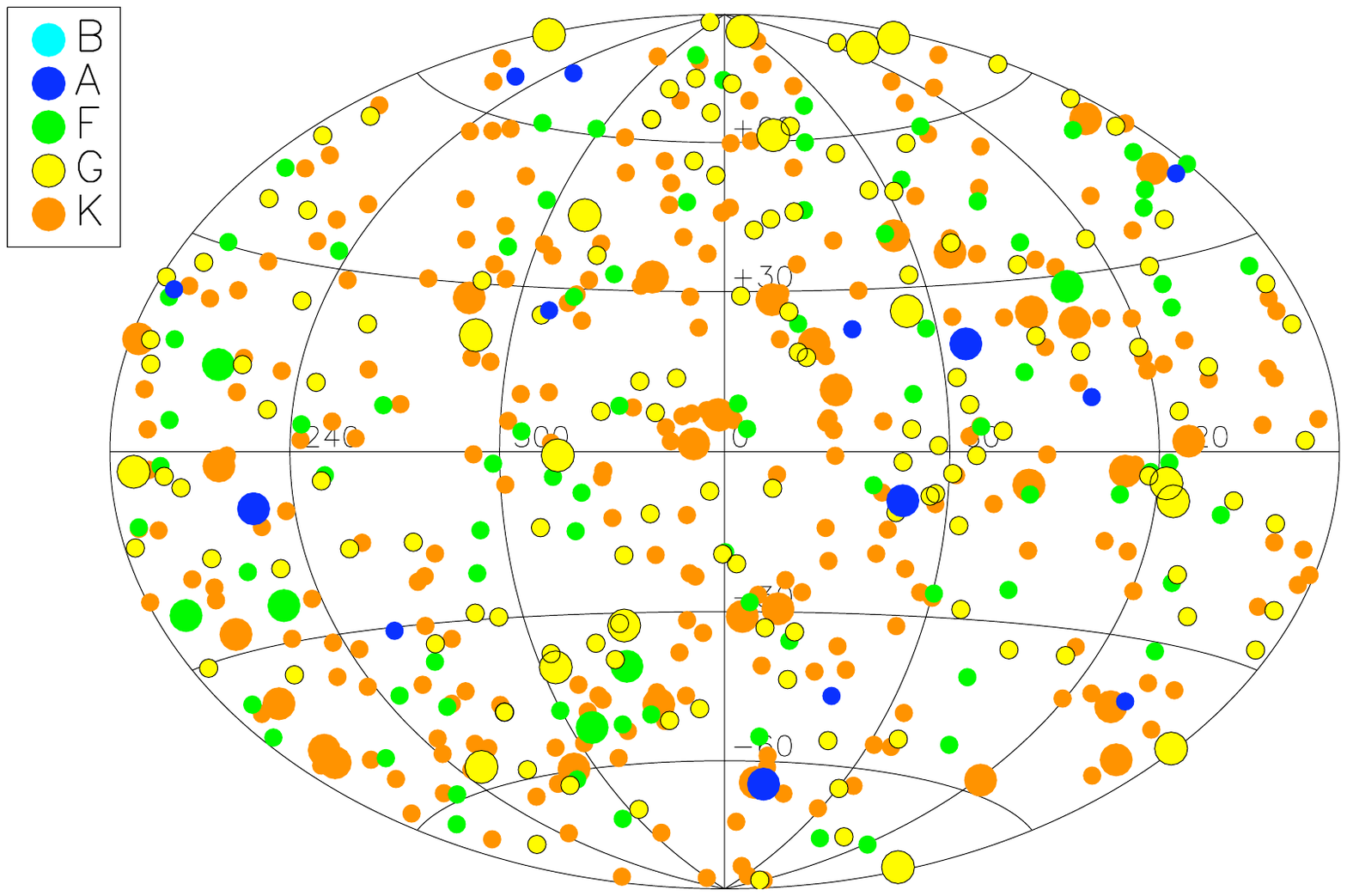}
\vspace*{-4.7in}
\flushright
\includegraphics[scale=0.4]{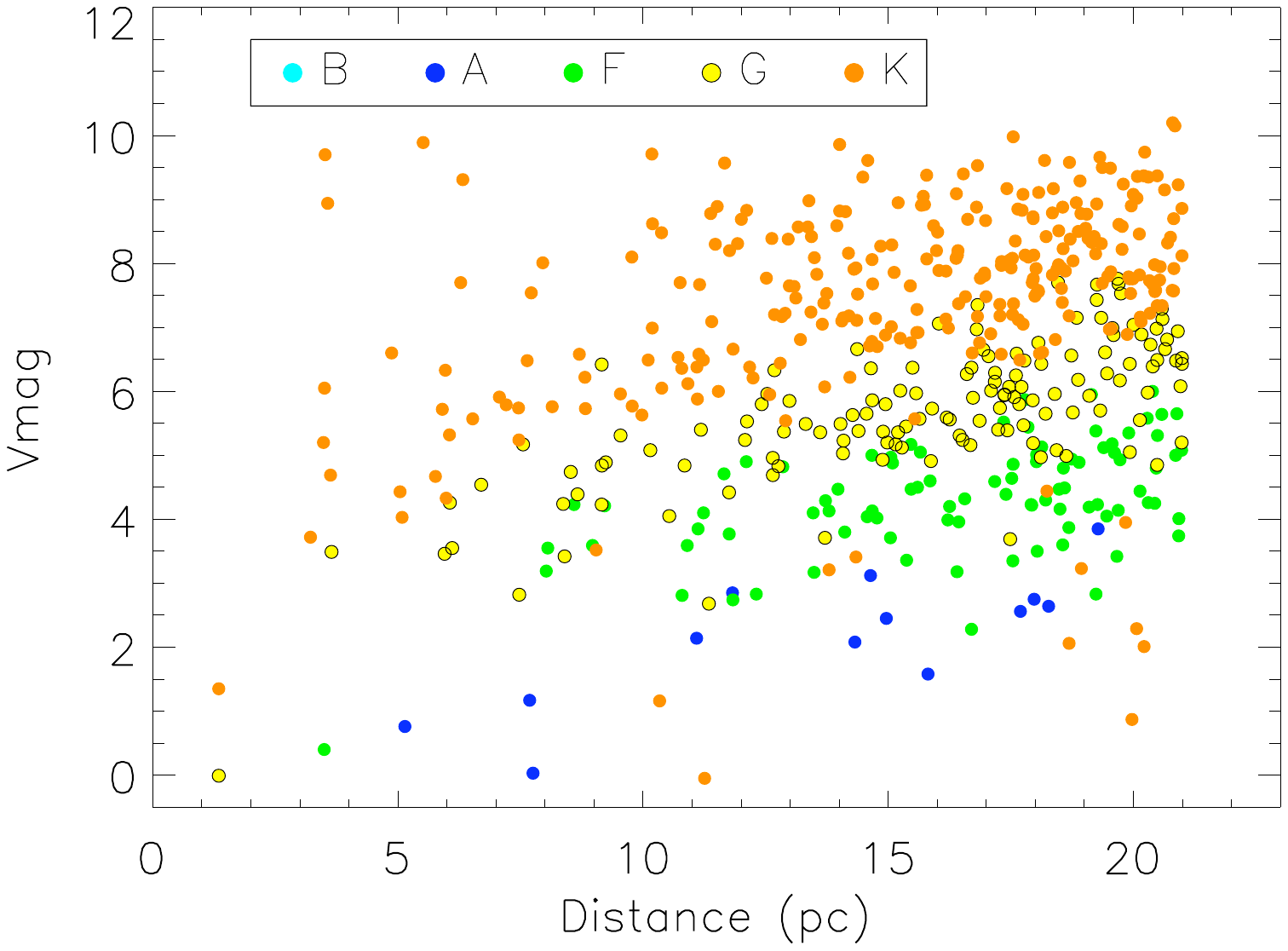}
\end{center}
\vspace*{-1.8in}
\caption{ \small Left: Spatial distribution of B, A, F, G, and K stars
with 20 pc of the Sun.  The spectral types are color coded.  Right:
Distribution of same set of stars over distance.  All stars are in the
Hipparcos catalog.
\label{fig:stars}}
\end{figure}

\ppar We do not understand the galactic environment of the Sun.
Interstellar clouds form the cosmic "ecosystem" through which the Sun
moves.  The only way that the physical properties of nearby
interstellar material (ISM) can be { \it surveyed} is through
observations of interstellar absorption lines in the ultraviolet and
far ultraviolet regions.  Over 500 B, A, F, G, and K stars within 20
pc are brighter than 9th mag, and 225 brighter than 6th mag
(Fig. \ref{fig:stars}).  In addition, at least two white dwarf stars
are in the upwind direction and within 20 pc of the Sun, with
V$\sim$11 mag.  Many of these stars could be used to survey the nearby
interstellar medium.  The key cloud parameters that need to be
measured are the cloud density, ionization, temperature and velocity.
Each of these variables has a significant effect on the heliosphere
boundary conditions.

A survey-mode study of nearby ISM is required to understand the solar
environment because most available target stars for this program are
cool and faint, and high S/N is needed.  An optimal survey would use
both STIS E140H and E230H data, at S/N$\ge 50$, for $\sim 100$ stars,
yielding data on Mg II, Mg I, H I or D I, C I, and S II, Zn II, and
where possible C II, and C II* and others.  Interesting regions
consist of those where there are no current data, where the velocities
of different clouds degenerate, where there currently is some evidence
but no hard data on an ionization gradient, or where an expected
velocity component is not seen.  In addition, some stars that have
already been observed by HST should be reobserved, in search of
time-variation in the hydrogen wall that would show that the solar
magnetic activity cycle impacts the hydrogen wall, an important
signature of the ISM-heliosphere interaction.

The minimum data set per star would include the 2800 A Mg II and Mg I
lines, from which the interstellar electron density can be derived for
both warm and cool clouds because of the high dielectronic
recombination rate of Mg II in warm $T > 6000$ K clouds.  Additional
information on either H I, D I, and an element with first ionization
potential (FIP) less than 13.6 eV such as sulfur or zinc, then allows
reconstruction of the ionization of the sightline.  Both the
$Copernicus$ and HST satellites have successfully observed these
species towards both warm and cool stars, however to acquire a data
set of consistent quality at high resolution and high-S/N requires a
large number of orbits to observe faint stars.  STIS MAMA count rates
limit possible observations for A and B stars, so that the grating
settings will vary between stars.  An example of an A-type target star
observed at high resolution and near- and far-UV, would require 2--8
orbits per star, for S/N$>100$.  Another example of a star for this
program is an 11th magnitude DA white dwarf star, which would require
medium resolution data, including both long and short wavelengths, for
5-6 orbits per star.

A survey program that consists of $\sim 100$ stars will require a
large number of orbits such as would be possible under a Multi-Cycle
Treasury program.  The second reason for advancing a survey of local
ISM as a MCTP program is that understanding our galactic environment
is a high-impact program, with goals that are difficult to achieve
piecemeal through small proposals.

%%%%%%%%%%%%%%%%%%%%%%%%
%% \input{fig_stars}

%%%%%%%%%%%%%%%%%%%%%%%%

\section{Interstellar Clouds and the Solar Galactic Environment}

\begin{figure}[!t]
\begin{center}
\flushleft
\includegraphics[scale=0.3]{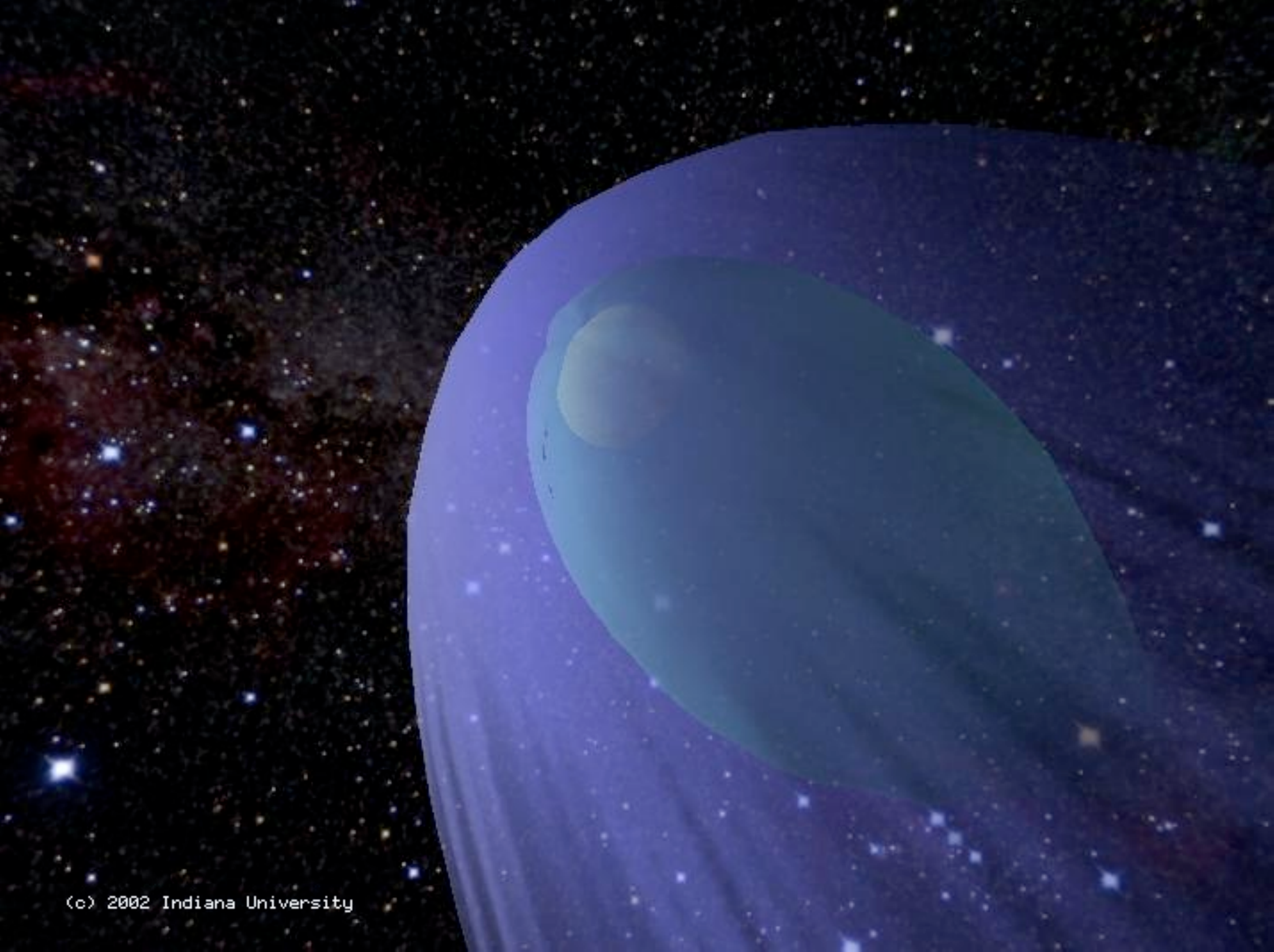}
\vspace*{-2.2in}
\flushright
\includegraphics[scale=0.4]{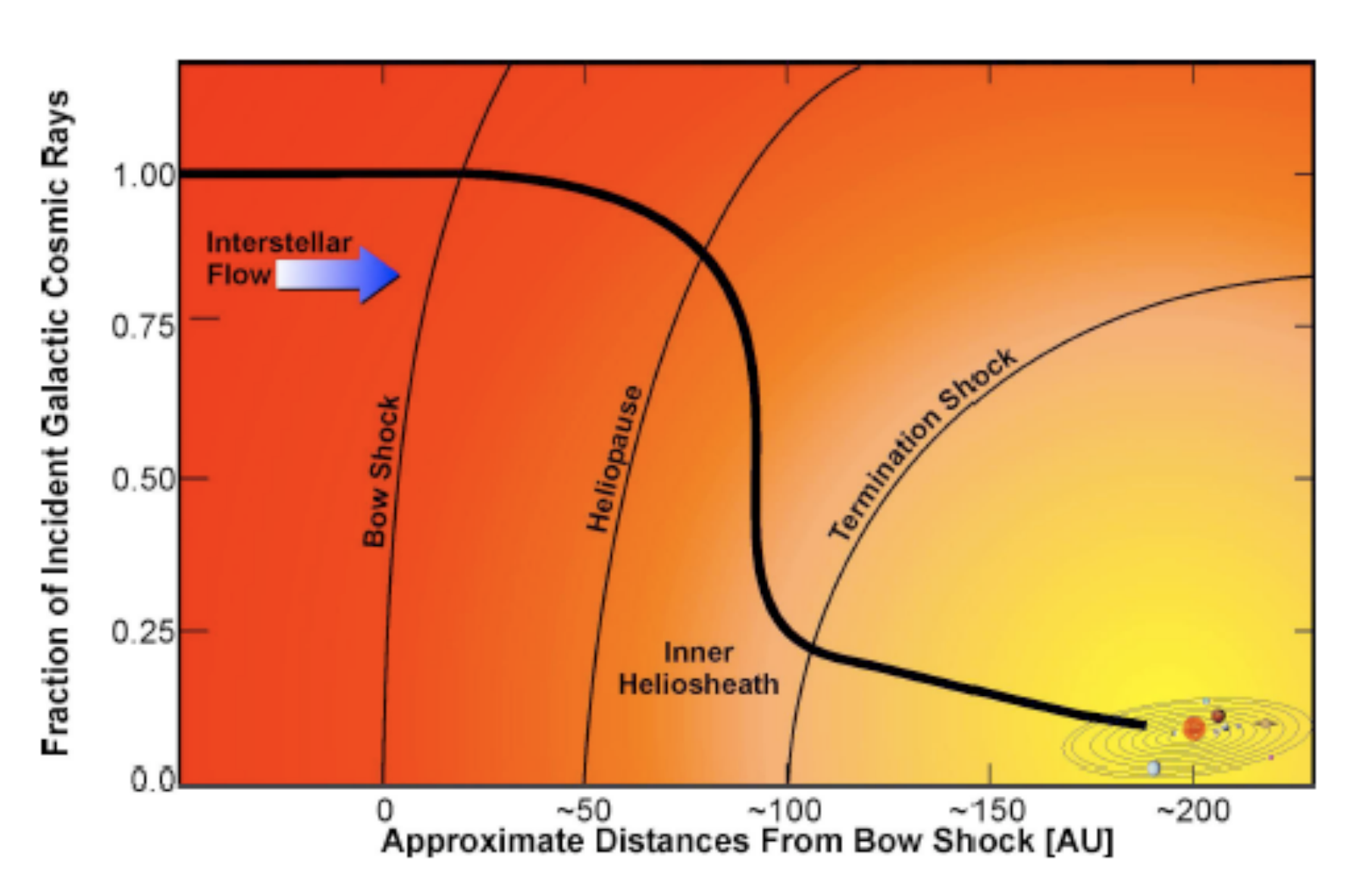}
\end{center}
\caption{ \small Left: The heliosphere today, with a weak Mach 1 bow
shock (purple), heliopause, termination shock (greenish semispherical
surface), and heliotail (figure from
http://antwrp.gsfc.nasa.gov/apod/ap020624.html ).  Right: The
importance of the configuration of the outer heliosphere for galactic
cosmic ray modulation, can be seen, since today most of the GCR
modulation is in the inner heliosheath whose properties are strongly
affected by the surrounding ISM (figure from Wimmer-Schweingruber,
McNutt, et al. 2007, submitted to \em{Experimental Astronomy}).
\label{fig:helio}}
\end{figure}

\ppar The Sun left the deepest void of the Local Bubble sometime
within the past 50,000--120,000 years and entered an interstellar wind
with an upwind direction towards the galactic center region.  The ISM
now surrounding the Sun is warm, partially ionized, and very low
density, $n_\mathrm{H} \sim 0.24$, H$^+$/H$_\mathrm{total} \sim 0.2$,
however the velocity of the ISM measured {\it inside} of the solar
system is not seen towards nearby stars in the upwind direction.  This
and other indicators suggest the Sun crosses into different clouds on
the order of $\sim 1/40,000$ years locally.  Over much shorter
timescales the Sun travels through strong ionization gradients in the
ISM, because of both the low cloud opacities and the distribution of
ionization sources.  The ISM surrounding the Sun shapes the
heliosphere, which in turn modulates the galactic cosmic ray flux at
Earth, see Fig. \ref{fig:helio}.  The cosmic ray record is traced by
the radioisotope record on Earth, which now provides coverage for the
past $\sim 0.5 $ Myrs, during which time the Sun has left the Local
Bubble interior and moved $\sim 8 - 10$ pc through space, and possibly
encountered different types of ISM.  An accurate map of the
distribution of nearby ISM, along with the physical properties of that
ISM, potentially will allow the history of the Earth's recent exposure
to galactic cosmic rays to be calculated from physical properties of
clouds that have crossed our cosmic trail in space.

The present day interaction between the Sun and surrounding ISM is
traced by: (1) Direct observations of ISM inside of the heliosphere,
including H, He, N, O, Ne, Ar, and interstellar dust grains.
Observations of the florescence of solar \lya\ radiation off of
interstellar \HI\ inside of the heliosphere by $Copernicus$ during the
mid-1970s, and HST during the mid-1990's, show that the Sun has been
in the same interstellar cloud for the past $\sim 20$ years, during
which time the heliosphere has sampled a pathlength of $\sim 100 AU$
of this cloud. (2) Lyman$\alpha$ observations of the "hydrogen wall",
formed where interstellar protons and hydrogen-atoms couple by charge
exchange as the ions are compressed and diverted around the
heliosphere.  (3) Energetic neutral atoms (ENAs) formed by
charge-exchange between interstellar \HI\ atoms and the solar wind.
The Interstellar Boundary EXplorer (IBEX) mission to be launched in
2008 will directly measure the properties of the ISM-heliosphere
interaction through measurements of ENAs formed in the outer
heliosphere.

%%%%%%%%%%%%%%%%%%%%%%%%
%%\input{fig_gcr}

%%%%%%%%%%%%%%%%%%%%%%%%

\section{The Heliosphere and Terrestrial Climate}

\ppar Interstellar clouds form the boundary conditions of the
heliosphere.  There are major differences in the way two hypothetical
interstellar clouds with the same heliocentric velocities, \Vhc, same
temperature and total density, but different ionizations, will
interact with the heliosphere.  This is because interstellar neutrals
enter the heliosphere, become ionized, and mass-load the solar wind,
thereby enhancing the solar wind barrier to galactic cosmic rays,
whereas in contrast interstellar ions increase the external ram
pressure of the ISM on the heliosphere.  The net result is that the
flux of galactic cosmic rays at Earth varies with the solar galactic
environment.  The cosmic ray attenuation by the heliosphere is
illustrated in Fig. \ref{fig:helio}.  There is evidence that galactic
cosmic rays seed the formation of terrestrial clouds, thereby cooling
Earth.\footnote{A concise reference describing the physical processes
that link the galactic environment of the Sun with the terrestrial
climate is the book {\it Solar Journey: The Significance of Our
Galactic Environment for the Heliosphere and Earth}, Ed. P. C. Frisch,
Springer 2006.}  There is strong circumstantial evidence that
increased fluxes of galactic cosmic rays at Earth (as measured by
$^{10}$Be isotopes in ice cores) are correlated with cold climate
spells, such as the Maunder Minimum $\sim 1640-1700$.

\clearpage
\newpage

\section{Appendix:  End Notes}

\ppar The above summary was prepared in response to a request for
3-page white papers discussing Multi-Cycle Treasury Programs for the
Hubble Space Telescope.  A more complete list of references for the
items listed above are given in this end note.  Fig. 1: The stars that
are plotted are from the Hipparcos star catalog
\cite{Perrymanetal:1997}.  Credits for the magnetohydrodynamic
heliosphere model visualization in Fig. 2, left, are A. Hanson, P. Fu,
P. Frisch, and \cite{Linde:1998}.  An additional reference for Fig. 2,
right, is \cite{Schwadronetal:2007}.  Discussions of the changing
galactic environment of the Sun over short time scales of $\sim 10^5$
years can be found in \cite{FrischSlavin:2006astra}.  Additional
information about the IBEX mission can be found in
\cite{McComasetal:2004} and at the URL http://www.ibex.swri.edu.  The
ionization gradient in the cloud surrounding the Sun, and constancy of
the cloud's velocity over $\sim 20$ years, are discussed in
\cite{SlavinFrisch:2007}; this reference also cites references for
measurements of interstellar H, He, O, N, Ne, and Ar inside of the
heliosphere.  Hydrogen-wall data are discussed in
\cite{Woodetal:2005}.  The processes that relate our galactic
environment to the terrestrial climate are discussed in chapters of the
book $Solar~Journey$, listed in footnote 2.

%\bibliography{frisch}
%\bibliographystyle{plain}

\end{document}